\documentclass[journal]{IEEEtran}
\usepackage{cite}

\ifCLASSINFOpdf
  \usepackage[pdftex]{graphicx}
\else
\fi
\usepackage{color}

\usepackage[cmex10]{amsmath}
\usepackage{amsthm}
\usepackage{amssymb}
\usepackage{array}

\ifCLASSOPTIONcompsoc
  \usepackage[caption=false,font=normalsize,labelfont=sf,textfont=sf]{subfig}
\else
  \usepackage[caption=false,font=footnotesize]{subfig}
\fi

\usepackage{multirow}
\usepackage{threeparttable}

\newtheorem{prop}{Proposition}

\begin{document}
%
\title{Density Evolution for Min-Sum Decoding of LDPC Codes Under Unreliable Message Storage}
%
%
%

\author{\authorblockN{Alexios Balatsoukas-Stimming,~\IEEEmembership{Student Member, IEEE} and Andreas Burg,~\IEEEmembership{Member, IEEE}}}
\maketitle

\begin{abstract}
We analyze the performance of quantized min-sum decoding of low-density parity-check codes under unreliable message storage. To this end, we introduce a simple bit-level error model and show that decoder symmetry is preserved under this model. Subsequently, we formulate the corresponding density evolution equations to predict the average bit error probability in the limit of infinite blocklength. We present numerical threshold results and we show that using more quantization bits is not always beneficial in the context of faulty decoders.
\end{abstract}


%
\IEEEpeerreviewmaketitle

\section{Introduction}


\IEEEPARstart{R}{ecently}, there has been strong interest in studying the performance of low-density parity-check (LDPC) codes under faulty decoding, which is motivated by the fact that VLSI integration has reached a level where it is difficult to guarantee fully reliable operation~\cite{Ghosh2010}. Specifically, in \cite{Varshney2011} the Gallager~A and the sum-product algorithms are analyzed under faulty decoding and an important concentration result is proved that makes the density evolution \cite{Richardson2001} analysis meaningful. Similar analyses are provided in \cite{Yazdi2013,Leduc-Primeau2012} for the Gallager B algorithm, while more general finite-alphabet decoders were considered in \cite{Huang2013}. A study of min-sum (MS) decoding~\cite{Wiberg1996}, which is widely used in practice, under faulty computations due to unreliable logic is presented in \cite{Kameni2013}.

In this work, we study MS decoding that is faulty due to unreliable memory. The reason for this approach is twofold. First, memory elements and flip-flops have been shown to be the first point of failure in digital circuits \cite{Chandrakasan2010}. Secondly, memory is the largest part, in terms of area, of most hardware LDPC decoders (see, e.g., \cite{Studer2008}). Thus, it is reasonable to assume that the remaining part of the decoder, which consists of logic, can be protected by standard fault-tolerance methods with small overhead. To this end, we introduce a bit-level fault model for unreliable memory reads and we prove that decoder symmetry is preserved under the aforementioned fault model. This fault model allows us to move away from the unrealistic assumption of uniformly distributed errors, which is found in all previous work.

\section{Quantized Min-Sum Decoding}\label{sec:ms}

\subsection{LDPC Codes and Channel Model}
An LDPC code $\mathcal{C}$ of blocklength $N$ can be defined as $\mathcal{C} \triangleq \left\{\mathbf{c} \in \{0,1\}^N: \mathbf{H}\mathbf{c} = \mathbf{0}\right\},$ where $\mathbf{H}$ is a sparse matrix with $H_{ij} \in \{0,1\}$ and operations are performed over GF$(2)$. $\mathbf{H}$ acts as an adjacency matrix for a bipartite graph which contains \emph{variable nodes} and \emph{check nodes}. Variable node $i$ is connected to check node $j$ iff $H_{ji}=1$. If all variable nodes have degree $d_v$ and all check nodes have degree $d_c$, then we say that the code is a $(d_v,d_c)$-regular code and that it belongs to the $(d_v,d_c)$-regular code \emph{ensemble}~\cite{Richardson2001}.
 
Transmission of $\mathbf{c} \in \mathcal{C}$ over an additive white Gaussian noise (AWGN) channel using binary phase-shift keying (BPSK) modulation is modeled as
\begin{align}
	y_i	& = x_i + w_i,~w_i \sim \mathcal{N}\left(0,\sigma ^2\right),~i = 1,\hdots,N,
\end{align}
where $x_i = 1 - 2c_i$. 
Let $L(y_i) \triangleq \ln \frac{p(y_i|x_{i}={+1})}{p(y_i|x_{i}={-1})}$ denote the channel log-likelihood ratio (LLR). The AWGN channel is output-symmetric in the sense that 
\begin{align}
	L(-y_i) & = -L(y_i). \label{eqn:chansym}
\end{align}

\subsection{Min-Sum Decoding}
In MS decoding, the message sent from variable node $i$ to check node $j$ is calculated as
\begin{align}
	m^{vc} _{i \rightarrow j}	& = m_i^0 + \sum _{j' \in \partial i / j} m^{cv}_{j' \rightarrow i}, \label{eqn:vc}
\end{align}
where $m ^0 _i = L(y_i) = y_i/2\sigma ^2$ and $\partial i/j$ denotes the set of all neighboring nodes of node $i$ except node $j$. The message sent from check node $j$ back to variable node $i$ is given by
\begin{align}
	m^{cv} _{j \rightarrow i}	& = \left(\prod _{i' \in \partial j / i} \text{sign}\left(m^{vc}_{i' \rightarrow j}\right)\right) \min _{i' \in \partial j / i} \left|m^{vc}_{i' \rightarrow j}\right|. \label{eqn:cv}
\end{align}
Decisions for codeword symbol $i$ are taken according to
\begin{align}
	\hat{x}_i	& = \text{sign} \left(m^0_i + \sum _{j' \in \partial i} m^{cv}_{j' \rightarrow i}\right). \label{eqn:dec}
\end{align}
Let $\Phi(m_0,m_1,\hdots,m_{d_v-1})$ and $\Psi(m_1,\hdots,m_{d_c-1})$ denote the update rules in (\ref{eqn:vc}) and (\ref{eqn:cv}), respectively. Let $b_i \in \{\pm 1\},~i=1,\hdots,d_c-1$. The following symmetries hold
\begin{align}
	\Phi(-m_0,-m_1,\hdots)	& = -\Phi(m_0,m_1,\hdots), \label{eqn:symvc} \\
	\Psi(b_1m_1,b_2m_2\hdots)	& = \prod _{i=1}^{d_c-1}b_i\Psi(m_1,m_2\hdots). \label{eqn:symcv}
\end{align}

\subsection{Message Quantization}
We assume uniform $b$-bit symmetric message quantization. In LDPC decoder implementations, $4 \leq b \leq 7$ are common values (see, e.g., \cite{Studer2008} and references therein). The quantizer is defined by
\begin{align}
	q(x) & = \text{sign}(x)\Delta \left\lfloor \frac{|x|}{\Delta} + \frac{1}{2}\right\rfloor,
\end{align}
where $\Delta$ denotes the quantization step. 
Let the set of all quantization levels be denoted by $\mathcal{L}$, so that $q(x) \in \mathcal{L},~\forall x \in \mathbb{R}$. The corresponding quantization intervals are
\begin{align}
	t_i = \left(\frac{l_{i-1}+l_i}{2},\frac{l_{i}+l_{i+1}}{2}\right],~i=0,\hdots,2^{b}-2,
\end{align}
where $l_{-1} = -\infty$ and $l_{2^{b}-1} = +\infty$. Results of (\ref{eqn:vc})--(\ref{eqn:dec}) that are smaller than $l_0$ or larger than $l_{2^b-2}$ are saturated to $l_0$ and $l_{2^b-2}$, respectively.

\section{Fault Model and Restriction to the All-One BPSK Codeword}\label{sec:error}
\subsection{Fault Model}
\begin{figure}[t]
	\centering
	\includegraphics[width=0.34\textwidth]{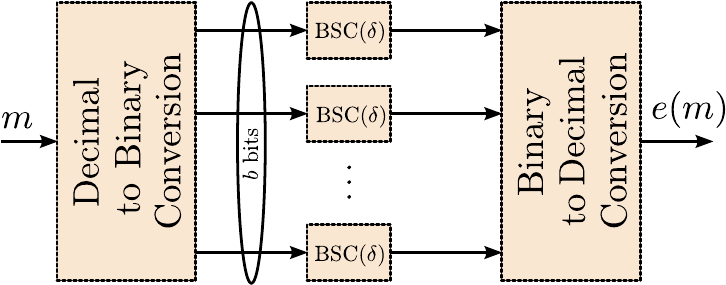}
	\caption{Message fault model: an incoming $b$-bit noiseless message of value $m$ is passed through $b$ independent BSC$(\delta)$ channels, resulting in the faulty message $e(m)$.}\label{fig:faultModel}
\end{figure}

\begin{figure}[t]
	\centering
	\subfloat[][]{\includegraphics[width=0.22\textwidth]{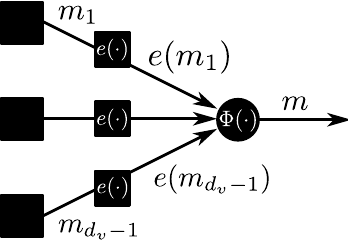}}$\quad$
	\subfloat[][]{\includegraphics[width=0.22\textwidth]{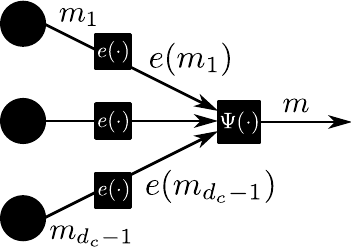}}
	\caption{Application of fault model to (a) variable node and (b) check node update rules. The fault model is a specific instance of the generic \emph{message wires} defined in \cite{Varshney2011}.}\label{fig:faultModelVarCheck}
\end{figure}

We assume sign-magnitude binary representation for all message values $m \in \mathcal{L}$. The memory read errors are modeled as independent and identically distributed (i.i.d.) random bit-flips. We assume that the memory errors are caused by single-event upsets, to which memory that is produced at very small lithographic nodes is particularly sensitive. Thus, all faults are transient, as in \cite{Varshney2011}, and independent of the stored message. More precisely, at each iteration, each bit of the binary representation of the messages used to compute \eqref{eqn:vc}--\eqref{eqn:dec} is passed through a binary symmetric channel (BSC) with crossover probability $\delta$, denoted by $\text{BSC}(\delta)$. We denote the set of all possible binary error patterns by $\mathcal{E}$ and the resulting faulty message after applying $e \in \mathcal{E}$ to a message of value $m$ by $e(m)$. The distribution of the error patterns is
\begin{align}
	\mathbb{P}(e)	& = \delta ^{\text{w}_{\text{H}}(e)}\left(1-\delta\right)^{b-\text{w}_{\text{H}}(e)},\quad e \in \mathcal{E}, \label{eqn:errordist}
\end{align}
where $\text{w}_{\text{H}}(e)$ denotes the Hamming weight of $e$. The fault model and its application to MS decoding are illustrated in Fig.~\ref{fig:faultModel} and Fig.~\ref{fig:faultModelVarCheck}, respectively.

\subsection{Restriction to the All-One BPSK Codeword}
Under the update rule symmetry defined in (\ref{eqn:symvc}) and (\ref{eqn:symcv}) and under channel symmetry, as defined in (\ref{eqn:chansym}), the probability of bit error is independent of the transmitted codeword~\cite{Richardson2001}. Thus, the asymptotic analysis of MS decoding can be restricted to the all-one BPSK codeword. The following proposition ensures that the same simplification can be applied to faulty quantized MS decoding with our error model.

\begin{prop}
	When messages are represented in sign-magnitude form, MS decoder symmetry is preserved under faulty decoding with read errors modeled as i.i.d. bit-flips.\label{lem:sym}
\end{prop}

\IEEEproof{
Due to quantizer symmetry, we have $q\left(L(-y_i)\right) = q(-L(y_i)) = -q(L(y_i)),$ so channel symmetry holds. Moreover, when using sign-magnitude representation where ``$+0$'' and ``$-0$'' exist as distinct values, it holds that 
\begin{align}
	e(-m) = -e(m),\quad \forall e \in \mathcal{E},~m \in \mathcal{L}.
\end{align} 
Thus, for the variable node update rule, we have
\begin{align}
	\Phi(e(-m_0),e(-m_1),\hdots)	& = \Phi(-e(m_0),-e(m_1),\hdots) \\
					& = -\Phi(e(m_0),e(m_1),\hdots). \label{eqn:symvcf}
\end{align}
Similarly, for $b_i \in \{\pm 1\},~i=1,\hdots,d_c-1,$ we have
\begin{align}
	\Psi(e(b_1m_1),e(b_2m_2),\hdots)& = \Psi(b_1e(m_1),b_2e(m_2),\hdots) \\
					& = \prod _{i=1}^{d_c-1}b_i\Psi(e(m_1),e(m_2),\hdots), \label{eqn:symcvf}
\end{align}
meaning that update rule symmetry holds for both variable nodes and check nodes. Moreover, we assume that, whenever $m=0$ appears, a uniform random choice between ``$+0$'' and ``$-0$'' is made, so that the bit error rate when $m=0$ is always $1/2$ independently of the codeword bit value.
}

\begin{figure*}
\begin{align}
	\Phi _+ ^{\ell}(m) & = \sum _{k = 0, \atop k\text{ even}}^{d_c-1} {d_c-1 \choose k}\left(A _+ ^{\ell}(m)\right)^{k}\left(A _- ^{\ell}(m)\right)^{d_c-k-1}, \qquad	\Phi _- ^{\ell}(m) = \sum _{k =0, \atop k\text{ odd}}^{d_c-1} {d_c-1 \choose k}\left(A _+ ^{\ell}(m)\right)^{k}\left(A _- ^{\ell}(m)\right)^{d_c-k-1} \label{eqn:phis}
\end{align}
\hrule
\end{figure*}

\section{Density Evolution}\label{sec:de}
Density evolution (DE) tracks the average probability density functions of the messages exchanged between the variable and check nodes at each decoding iteration in the limit of infinite blocklength~\cite{Richardson2001}. DE operates under the assumption that all messages are independent because it can be shown that the decoding graph is asymptotically cycle-free. A concentration result guarantees that the performance of individual codes chosen from an ensemble is close to the ensemble average performance with high probability~\cite{Richardson2001}.

The existence of transient errors using the error model introduced in Section~\ref{sec:error} does not affect the asymptotic cycle-free property of the decoding graph. However, care still needs to be taken to ensure that all messages involved in decoding are independent. With our error model the faulty messages are independent, because the corresponding non-faulty messages from which they are derived are independent and the errors affecting a specific message are independent of the message value. A concentration result that is similar to the concentration result for noiseless decoders was proved in \cite{Varshney2011}.

\subsection{Density Evolution for Quantized MS Decoding}
Let $p^{\ell}(m)$ and $q^{\ell}(m)$ denote the probability mass functions (PMFs) of the variable-to-check and the check-to-variable messages at iteration $\ell \geq 1$, respectively, and let $p^{0}(m)$ denote the PMF of the channel LLR messages assuming that the all-one BPSK codeword was transmitted. We have
\begin{align}
	p^0\left(l_i\right)	& = \frac{1}{\sqrt{8\pi \sigma ^{-2}}}\int _{t_i} e^{-\left(x-\frac{2}{\sigma ^2}\right)^2 \cdot \frac{\sigma^2}{4}} dx,\quad \forall l_i \in \mathcal{L}.
\end{align}
The check-to-variable message density is given by
\begin{align}
	q^{\ell}(m)	& = \left\{ 
		\begin{array}{ll}
			\Phi ^{\ell}_-(m) - \Phi ^{\ell}_-(m-1), & m < 0, \\
			1 - \left(1-p^{\ell}(0)\right)^{d_c-1}, & m = 0, \\
			\Phi ^{\ell}_+(m+1) - \Phi ^{\ell}_+(m), & m > 0,
		\end{array} \right.
		\label{eqn:decv}
\end{align}
where $\Phi ^{\ell}_-(m)$ and $\Phi ^{\ell}_+(m)$ are defined in \eqref{eqn:phis} and 
\begin{align}
	A _+ ^{\ell}(m) & = \sum _{x=m}^{l_{2^b-2}}p^{\ell}(x), \quad m > 0, \label{eqn:aplus} \\
	A _- ^{\ell}(m) & = \sum _{x=l_0}^{m}p^{\ell}(x), \quad m < 0. \label{eqn:aminus}
\end{align}
The variable-to-check message density is given by
\begin{align}
	p^{\ell}(m)	& = p^0(m) \otimes \left(q^{(\ell-1)}(m)\right)^{\otimes (d_v-1)}, \label{eqn:devc}
\end{align}
where $\otimes$ denotes the convolution and $q^0(m) = \delta [m]$, where $\delta [m]$ is the Kronecker delta function. The density of the quantity used for bit decisions is given by
\begin{align}
	d^{\ell}(m)	& = p^0(m) \otimes \left(q^{(\ell-1)}(m)\right)^{\otimes d_v}. \label{eqn:dedec}
\end{align}
When applying \eqref{eqn:devc} and \eqref{eqn:dedec}, any probability mass that corresponds to values smaller than $l_0$ or larger than $l_{2^b-2}$ is added to the mass corresponding to $l_0$ or $l_{2^b-2}$, respectively.

\subsection{Density Evolution for Faulty Quantized MS Decoding}

Let $f_{\delta}(P)(m)$ denote the probability of a faulty message $m,~m\in\mathcal{L}$, where $P$ is the distribution of the non-faulty messages $m'$, i.e., $P$ can be $p^{\ell}$ or $q^{\ell}$. We have
\begin{align}
	f_{\delta}(P)(m)	& = \sum _{\substack{e \in \mathcal{E},m'\in\mathcal{L}:\\e(m')=m}}P(m')\mathbb{P}(e). \label{eqn:disterr}
\end{align}
For each value $m$, there are $2^b$ pairs $(e,m')$ such that $e(m')=m$. Since there are $2^b-1$ values for $m$,\footnote{Recall that the decimal value $0$ corresponds to two binary patterns (i.e., "$+0$" and "$-0$").} evaluating $f_{\delta}(P)$ requires the calculation of approximately $2^{b+1}$ terms.

Unreliable memory reads cause errors in the \emph{input} messages of \eqref{eqn:vc}--\eqref{eqn:dec}. Thus, DE for faulty MS decoding with transient memory read errors can be formulated by replacing the $p^{\ell}$ and $q^{\ell}$ distributions that appear on the right-hand side of \eqref{eqn:decv}--\eqref{eqn:dedec} with $f_{\delta}\left(p^{\ell}\right)$ and $f_{\delta}\left(q^{\ell}\right)$, respectively.

\subsection{Bit Error Probability and Threshold}
Let $P_e^{\ell}(\sigma ^2)$ denote the bit error probability at iteration $\ell$ when transmission takes place over an AWGN channel with noise variance $\sigma ^2$, defined as
\begin{align}
	P_e^{\ell}(\sigma ^2) & \triangleq \frac{1}{2}d^{\ell}(0) + \sum _{m=l_0}^{l_{2^{b-1}-2}}d^{\ell}(m). \label{eqn:deerr}
\end{align} 
The threshold for faulty decoding is defined as~\cite{Varshney2011}
\begin{align}
	\sigma _*^2(\eta)	& \triangleq \sup \left\{ \sigma ^2\geq 0: \lim _{\ell \rightarrow \infty} P_e^{\ell}(\sigma ^2) \leq \eta \right\}. \label{eqn:threshredef}
\end{align}
For all examined $(d_v,d_c)$-regular ensembles and $\sigma ^2$, our numerical results show that $P_{e}^{\ell}\left(\sigma ^2\right) \geq \delta$. This observation can help us in choosing a meaningful value for $\eta$. Specifically, we choose $\eta = \alpha \delta$, for some $\alpha > 1$. If $\alpha$ is chosen so that $\alpha\delta$ lies within the waterfall region of the code, then the value of $\alpha$ does not have a significant effect on the computed threshold. To make the dependence on $\alpha$ and $\delta$ explicit, we denote the threshold by $\sigma _*^2(\alpha,\delta)$.



\section{Numerical Results}\label{sec:numerical}
\subsection{Decoding Threshold}\label{sec:numa}
We computed $\sigma _*^2(\alpha,\delta)$ under MS and faulty MS decoding for various $(d_v,d_c)$-regular ensembles of rate $0.5$ and for various values of $\delta$, with $\alpha = 10$ and $b = 5$ bits. The maximum number of iterations is set to $\ell _{\max} = 200$. Quantization is performed with $\Delta = 1$. For fair comparison, the definition in (\ref{eqn:threshredef}) was used for both MS and faulty MS decoding.

\begin{figure}[t]
	\centering
	\includegraphics[width=0.44\textwidth]{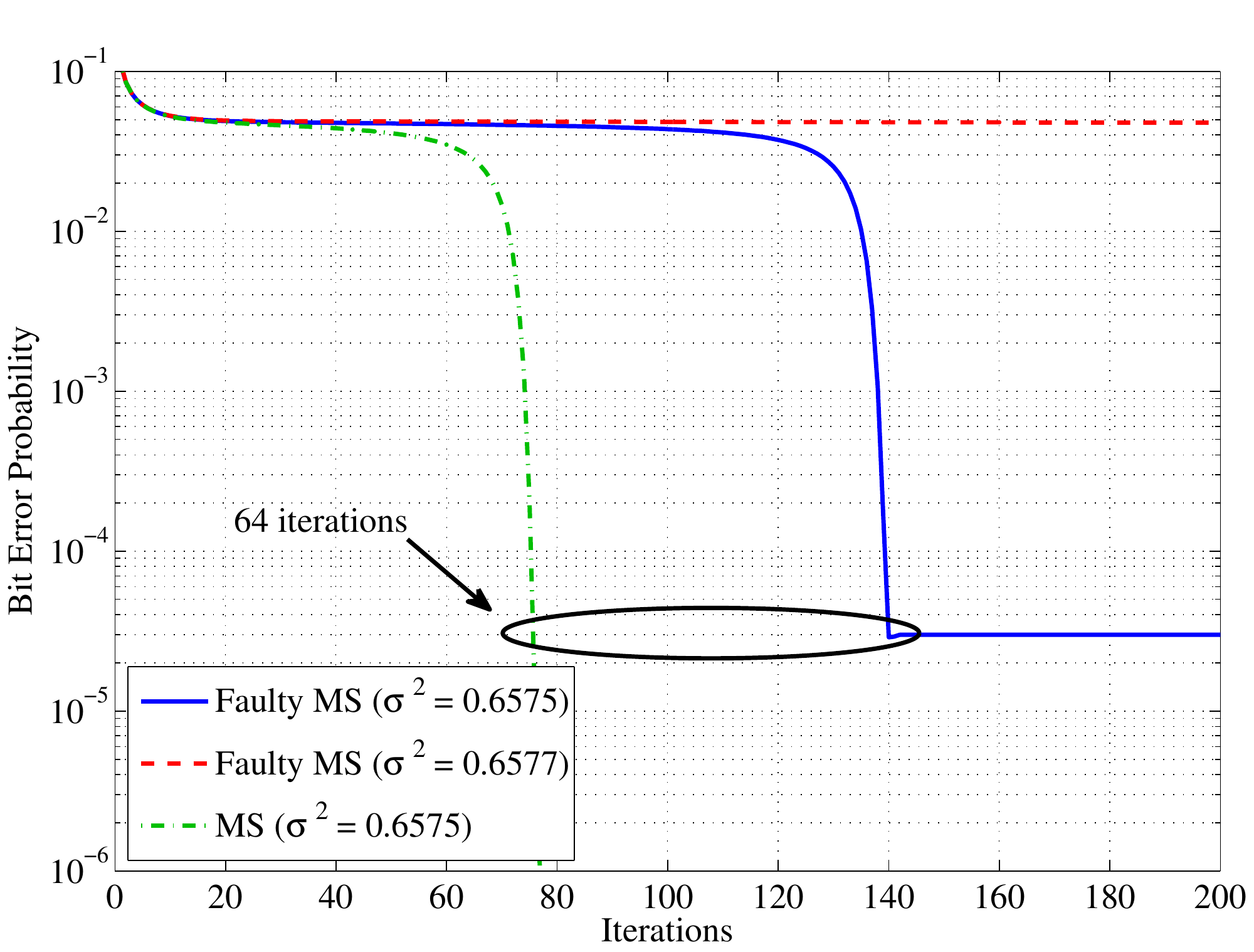}
	\caption{Error probability for a $(3,6)$-regular LDPC code under faulty MS and MS decoding for $\delta = 10^{-5}$.}\label{fig:faultyMSPe1e-5}
\end{figure}

\begin{figure}[t]
	\centering
	\includegraphics[width=0.44\textwidth]{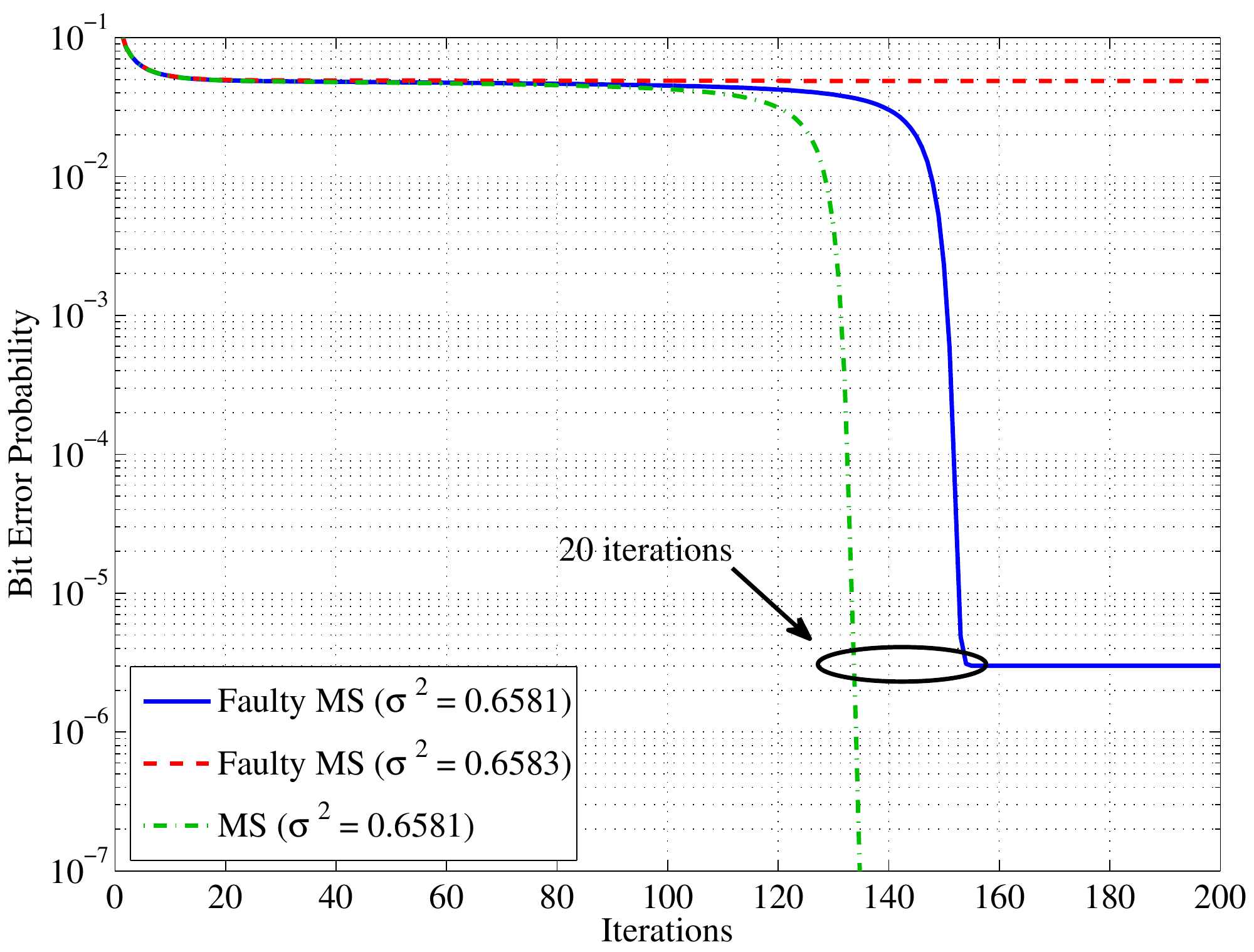}
	\caption{Error probability for a $(3,6)$-regular LDPC code under faulty MS and MS decoding for $\delta = 10^{-6}$.}\label{fig:faultyMSPe1e-6}
\end{figure}

The results are summarized in Table~\ref{tab:res}. The evolution of $P_e^{\ell}(\sigma ^2)$ as a function of $\ell$ for the $(3,6)$-regular ensemble and for two indicative cases of $\delta = 10^{-5}$ and $\delta=10^{-6}$ under MS and faulty MS decoding is presented in Fig.~\ref{fig:faultyMSPe1e-5} and Fig.~\ref{fig:faultyMSPe1e-6}, respectively. The error floor for faulty MS decoding is very apparent. This visualization also enables us to calculate the overhead, in terms of additional iterations, introduced by faulty decoding. In Fig.~\ref{fig:faultyMSPe1e-5}, the faulty MS decoder for $\delta = 10^{-5}$ requires $64$ more iterations than the MS decoder to achieve the same bit error probability when operating at the faulty MS decoder's threshold. In Fig.~\ref{fig:faultyMSPe1e-6}, we see that for a smaller $\delta$, the difference in iterations is smaller, as intuitively expected.

\begin{table}[t]
	\centering
	\caption{Thresholds of various $(d_v,d_c)$-regular codes under MS and faulty MS decoding for $\alpha=10$ and $b = 5$ bits.}\label{tab:res}
	\begin{tabular}{c|c|c|cccc}
		$(d_v,d_c)$		&	& $\delta$ 									& $10^{-3}$	& $10^{-4}$	& $10^{-5}$	& $10^{-6}$ \\
		\hline
		\multirow{2}{*}{(3,6)}	& MS	& $\sigma _*^2(\alpha,\delta)$	& 0.6579	& 0.6579	& 0.6579	& 0.6582 \\
					& F-MS 	& $\sigma _*^2(\alpha,\delta)$				& 0.5703	& 0.6518 	& 0.6576	& 0.6582 \\
		\hline
		\multirow{2}{*}{(4,8)}	& MS	& $\sigma _*^2(\alpha,\delta)$	& 0.5486	& 0.5486	& 0.5486	& 0.5486 \\
					& F-MS 	& $\sigma _*^2(\alpha,\delta)$				& 0.5077	& 0.5446	& 0.5482	& 0.5486 \\
	\hline
		\multirow{2}{*}{(5,10)}	& MS	& $\sigma _*^2(\alpha,\delta)$	& 0.4793	& 0.4793	& 0.4793	& 0.4793 \\
					& F-MS 	& $\sigma _*^2(\alpha,\delta)$				& 0.4473	& 0.4761	& 0.4790	& 0.4792 \\
	\hline
		\multirow{2}{*}{(6,12)}	& MS	& $\sigma _*^2(\alpha,\delta)$	& 0.4320	& 0.4320	& 0.4320	& 0.4320 \\
					& F-MS 	& $\sigma _*^2(\alpha,\delta)$				& 0.4041	& 0.4292	& 0.4317	& 0.4320 \\
	\end{tabular}
\end{table}

It is interesting to note that the threshold generally decreases when $d_v$ and $d_c$ are increased, but the resulting code ensembles seem to be more resilient to errors. The loss in $\sigma _*^2(\alpha,\delta)$ as $\delta$ is increased is smaller for larger $(d_v,d_c)$ pairs. This interesting trade-off between threshold and error resilience motivates the design of irregular LDPC codes for faulty MS decoding.

\subsection{Are More Bits Always Better?}
In faulty decoding it can not be claimed in advance that increasing the number of quantization bits $b$ will result in better performance, because by increasing $b$ we also increase the number of faults in the decoder. The additional bits can be used either to increase the dynamic range (DR) or to increase the precision (PR) of the messages. The DR case corresponds to quantization with a fixed step size, while in the PR case the quantization step is a function of $b$. 

\begin{figure} 
	\centering
	\includegraphics[width=0.44\textwidth]{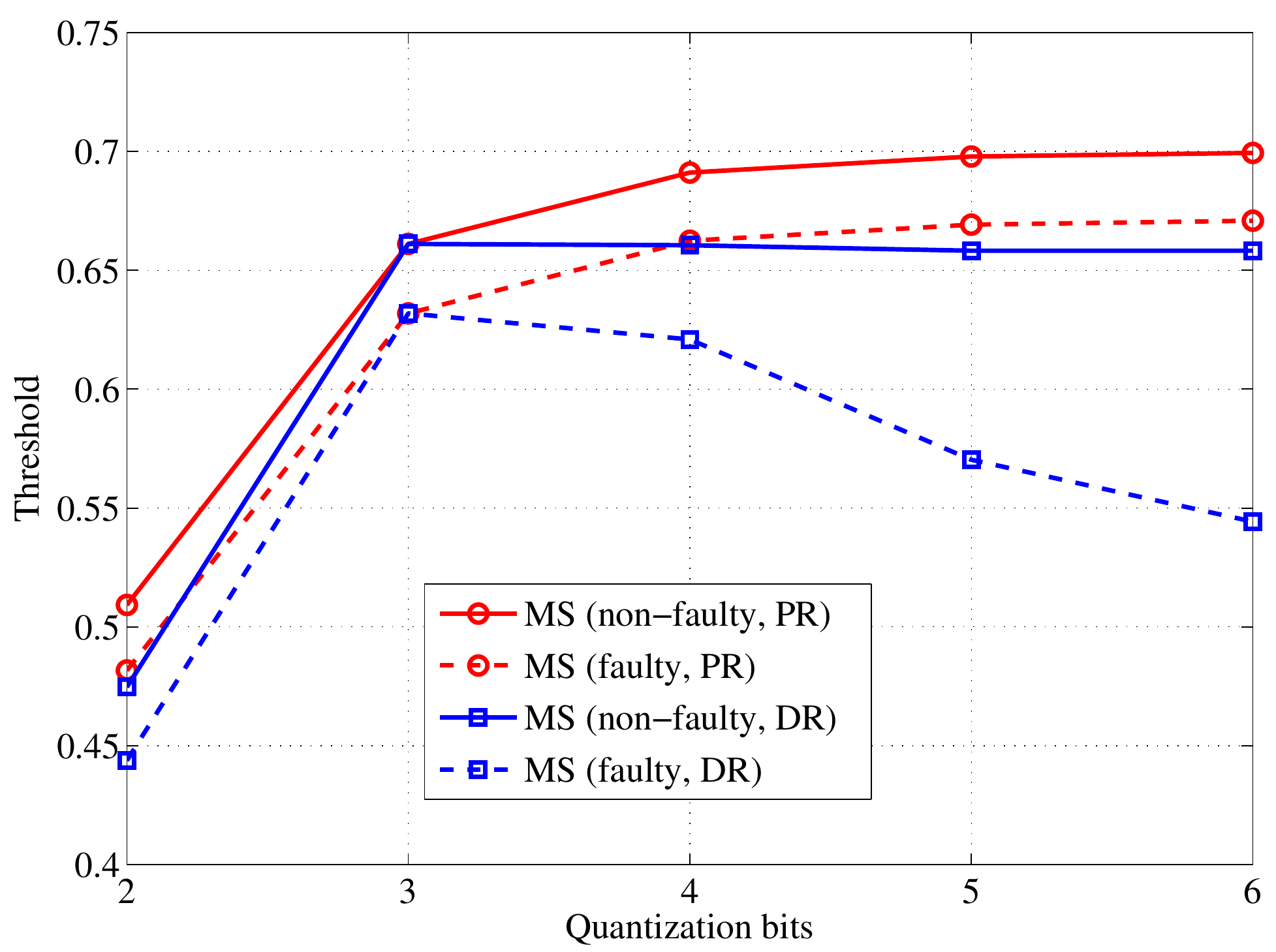}
	\caption{Error probability for a $(3,6)$-regular LDPC code under faulty MS and MS decoding for $\delta = 10^{-3}$ for different numbers of quantization bits.}\label{fig:threshVsBits}
\end{figure}

In Fig.~\ref{fig:threshVsBits} we present indicative threshold results for the DR case with $\Delta _{\text{DR}} = 1$, as used in Section~\ref{sec:numa}, and for the PR case with $\Delta _{\text{PR}} = 2^{3-b}$, which we empirically found to provide good performance with fault-free decoding in both cases, and $\delta = 10^{-3}$ and $\alpha = 10$. We observe that increasing the dynamic range does not offer any benefits after $b=3$ for fault-free decoding in the examined scenario, and that PR quantization provides better performance than DR quantization. More importantly, however, in the DR case the performance actually degrades for $b \geq 3$. This behavior can be explained intuitively as follows. In the DR case, bit-flips in the additional bits cause increasingly larger errors in the message values, whereas in the PR case these errors become smaller when $b$ is increased. 

We have demonstrated that, if one is not careful, the performance of faulty MS decoding can actually degrade with increasing $b$. Thus, we conclude that more bits are, perhaps counter-intuitively, not always better in the context of faulty MS decoding.

\ifCLASSOPTIONcaptionsoff
  \newpage
\fi



\bibliographystyle{IEEEtran}
\bibliography{bibliography.bib}
\end{document}